\documentclass
[twocolumn,secnumarabic,amsmath,amssymb,showpacs,nobibnotes,aps, prl]{revtex4-1}

\usepackage{amssymb}

\usepackage{graphicx}
\usepackage{dcolumn}
\usepackage{bm}
\usepackage{color}

\begin{document}
\title{Transport equation for plasmas in a stationary-homogeneous turbulence}
\author{Shaojie Wang}
\email{wangsj@ustc.edu.cn}
\affiliation{Department of Modern Physics, University of Science and Technology of China, Hefei, 230026, China}
\date{\today}

\begin{abstract}
For a plasma in a stationary homogeneous turbulence, the Fokker-Planck equation is derived from the nonlinear Vlasov equation by introducing the entropy principle. The ensemble average in evaluating the kinetic diffusion tensor, whose symmetry has been proved,  can be computed in a straightforward way when the fluctuating particle trajectories are provided. As an application, it has been shown that a mean electric filed can drive a particle flux through the Stokes-Einstein relation, independent of the details of turbulence.
\end{abstract}

\pacs{52.25.Dg, 52.25.Fi, 52.35.Ra}

\maketitle

\section{Introduction}
Fokker-Planck transport equation is widely used in scientific research fields, for examples but not limited to, the Brownian motion in statistical physics \cite{Einstein1905a}, the statistical dynamics of stellar systems in astronomy \cite{ChandrasekharRMP43}, the collisional dissipation of plasmas \cite{LifshitzBOOK81}, the quasilinear transport of space plasmas and fusion plasmas \cite{,BalescuBook05,ChenJGR99,KominisPRL10,WangPoP12}, the nonlinear turbulent transport of magnetic fusion plasmas \cite{BakerPoP98,EscandePRL07}, biologies \cite{CodlingInt08,SjobergCVS09}, and financial systems \cite{FriedrichPRL00, SornettePA01}. The Fokker-Planck equation for collisional dissipation is derived from the transition probability \cite{Einstein1905a,LifshitzBOOK81}; for quasilinear plasma transport, it is derived from the nonlinear Vlasov equation or the gyro-kinetic Vlasov equation \cite{FriemanPF82,BrizardRMP07} based on the perturbation theory, therefore its applicability is limited. The present Fokker-Planck equation for the nonlinear plasma turbulence is phenomenological or based on various reduced models \cite{BalescuBook05}, therefore, the full transport matrix predicted by these models can not be directly compared with the large-scale first-principle nonlinear turbulence simulations \cite{LinScience98,ChengJCP76,SonnendruckerJCP99}. Here we report a first-principle derivation of the Fokker-Planck equation for a stationary-homogeneous turbulence from the nonlinear Vlasov equation.

\section{Phase-space transport equation}
We begin with the nonlinear Vlasov equation
\begin{equation}
\partial_t f\left(\bm Z, t\right)+\partial_{\bm {Z}}\cdot\left[\dot{ \bm {\mathcal{Z} } }\left(\bm Z, t\right)f\right]=0,\label{eq:V}
\end{equation}
with $\bm Z$ the 6-dimensional phase space point, and $f$ the distribution function. When using the particle coordinate, $\bm Z=\left(\bm x, \bm v\right)$, with $\bm x$ and $\bm v$ the position and velocity of the particle, respectively; $\partial_{\bm Z}\cdot \bm A =\partial_{\bm x}\cdot \bm A^ {\bm x}+ \partial_{\bm v}\cdot \bm A^ {\bm v}$. When using the guiding-center coordinate in the gyro-kinetic theory, $\bm Z=\left(\bm X, v_{\parallel},\mu,\xi\right)$, which denotes the guiding-center position, the parallel velocity, the magnetic moment, and the gyro-angle, respectively; note that in the gyro-kinetic theory, since $d\mu/dt=0$ and $\partial_{\xi}f=0$ due to the gyro-symmetry, the system is reduced to a 4-dimensional one, and one has $\partial_{\bm Z}\cdot \bm A=\frac{1}{B_{\parallel}^{*}}\frac{\partial}{\partial \bm X}\cdot B_{\parallel}^{*} \bm A ^{\bm X}+\frac{1}{B_{\parallel}^{*}}\frac{\partial}{\partial v_{\parallel}}B_{\parallel}^{*} \bm A^{v_{\parallel}}$, with $B_{\parallel}^{*}=\bm b \cdot \left(\bm B +\frac{m}{e}v_{\parallel}\nabla\times \bm b\right)$, $\bm b=\bm B/B$; $m$ and $e$ are the mass and charge of the particle, respectively; $\bm B$ is the magnetic field.

It should be emphasized here that the phase-space velocity in Eq. (\ref{eq:V}), $\dot{\bm {\mathcal {Z} } }\left(\bm Z, t\right)$ is written explicitly in the Eulerian representation; it is given by the Hamiltonian dynamics, for example

\begin{equation}
\dot{\bm{\mathcal {Z} } }\left(\bm Z, t\right)=\left\{\bm Z,H_0+e\delta \Phi\right\},
\end{equation}
for the particle dynamics in the electrostatic turbulence, with $H_0=\frac{1}{2}m v^2$ the unperturbed Hamiltonian of the particle and $\delta \Phi$ the fluctuating electrostatic potential; $\{,\}$ is the Poisson bracket.

\begin{equation}
\dot{\bm{\mathcal {Z} } }\left(\bm Z, t\right)=\left\{\bm Z,H_0+ e\left\langle\delta \Phi\right\rangle_{gyro}\right\},
\end{equation}
for the gyro-center dynamics used in the gyro-kinetic Vlasov equation \cite{BrizardRMP07}; $H_0=\frac{1}{2}mv_{\parallel}^2+\mu B$, with $B$ the equilibrium magnetic field. $\left\langle\delta \Phi\right\rangle_{gyro}$ is the fluctuating potential averaged over the fast gyro-motion.

Note that the fluctuating electrostatic potential is associated with the fluctuating distribution function through the Poisson equation, therefore, the Vlasov-Poisson system is a nonlinear complex system.

The Vlasov or gyro-kinetic Vlasov equation can be solved by using the characteristic method,
\begin{equation}
f\left(\bm Z, t\right)=f\left[\bm{\mathcal {Z}}\left(t_0;\bm Z, t\right),t_0\right], \label{eq:cha}
\end{equation}
with the characteristic line or the phase-space trajectory given by integrating the Hamiltonian equations of motion,
\begin{equation}
\bm{ \mathcal{Z} }\left( t_0;\bm Z, t\right)=\int_{t}^{t_0}d_L t' \dot {\bm{\mathcal{Z} } }\left[\bm{\mathcal{Z} }\left(t';\bm Z, t\right), t'\right]+\bm Z,
\end{equation}
where we have used $d_L t'$ to emphasize that it is a "Lagrangian" integral; the argument in $\dot {\bm{\mathcal{Z} } }\left(\bm Z , t\right)$ should be understood as the Lagrangian variable. Clearly, $\bm{\mathcal{Z} }\left(t_0;\bm Z, t\right)$ denotes the phase space coordinate at time $t_0$ of a particle which passes through $\bm Z$ at time $t$.

To derive the transport equation, we shall make four basic assumptions for the plasma turbulence as follows.

(1) The turbulence is stationary and homogeneous.

(2) The characteristic scale length of the turbulent fluctuation, $\lambda_c$, is much smaller than the characteristic scale length of the macroscopic thermodynamic variables (such as the averaged density and pressure), $L$; $\epsilon\equiv \lambda_c/L\ll1$. The characteristic time scale of the turbulence, $\tau_c$, is much shorter than the characteristic time scale of the macroscopic observables, $\tau_E$. In detail, we shall assume that $\tau_c/\tau_E\sim \mathcal{O}\left(\epsilon^2\right)$, and by assuming that $\partial_t \delta f \sim \partial_t F$, we have $\delta f/F\sim \mathcal{O}\left(\epsilon^2\right)$ (here $\delta f$ and $F$ are the fluctuating part and the average part of distribution function, respectively).

(3) The ensemble-averaged distribution function that determines the entropy of the system can be found by taken the time-average on the time scale longer than $\tau_c$ but shorter than $\tau_E$, with $\bm Z$ the phase-space point fixed.

(4) The local entropy production rate is positive when the ensemble-averaged distribution is inhomogeneous. Since there is no present experimental observations inconsistent with this statement, it may be raised as a principle; hereafter, it may be referred to as the entropy principle, which can be taken as an extension of the second law of thermodynamics to a nonlinear turbulent system.

Assumption (3) indicates that the ensemble-averaged distribution function $F\left(\bm Z, t\right)$ can be written as
\begin{equation}
F\left(\bm Z, t\right)\equiv \left\langle f \right\rangle \left(\bm Z, t\right)= \lim_{T/\tau_c\rightarrow\infty}\frac{1}{T}\int_{t-\frac{T}{2}}^{t+\frac{T}{2}}d_E t'f\left(\bm Z, t'\right),
\end{equation}
where $d_E t'$ is used to emphasize that it is a "Eulerian" integral; $\bm Z$ in the kernel should be understood as a constant.

Clearly, one has $f=F+\delta f$, and $\left\langle \delta f\right\rangle=0$. Using Eq. (\ref{eq:cha}), one finds
\begin{equation}
f\left(\bm Z, t\right)=F\left[\bm{\mathcal{Z} }\left(0;\bm Z, t\right),0\right]+\delta f \left[\bm{\mathcal{Z} }\left(0;\bm Z, t\right),0\right].
\end{equation}

Following assumption (2), one can write the first term as
\begin{equation}
F\left(\bm Z, t\right)= F\left(\bm Z, 0\right)-\int_{0}^{t}d_L t' \dot{ \bm{\mathcal{Z} }}\left[\bm{\mathcal{Z} }\left(t';\bm Z, t\right), t'\right] \cdot \partial_{\bm Z} F\left(\bm Z, 0\right).
\end{equation}

Substituting the above solution into Eq. (\ref{eq:V}), and integrating the resulting equation with respect to $t$ from $-T/2$ to $T/2$ with $\bm Z$ fixed, for $\tau_E \gg T\gg\tau_c$ or in the limits $T/\tau_c\rightarrow \infty$ and $T/\tau_E \rightarrow 0$, one finds
\begin{equation}
\partial_t F\left(\bm Z, 0\right)+\partial_{\bm Z}\cdot \left[\left\langle\dot{\bm{\mathcal{Z} }}\right\rangle F \left(\bm Z, 0\right) - \bm D\cdot \partial_{\bm Z} F\left(\bm Z, 0\right)\right]=\Delta,\label{eq:FP0}
\end{equation}
with the phase-space diffusion tensor
\begin{equation}
\bm D \left(\bm Z\right)=\frac{1}{T}\int_{-\frac{T}{2}}^{\frac{T}{2}}d_E t \dot{\bm{\mathcal{Z} }}\left(\bm Z, t\right) \int_{0}^{t}d_L t' \dot{\bm{\mathcal{Z} }}\left[\bm{\mathcal{Z} }\left(t';\bm Z, t\right), t'\right],\label{eq:D0}
\end{equation}
and the residual term
\begin{equation}
\Delta=-\partial_{\bm Z}\cdot\left\{\left\langle\dot{\bm{\mathcal{Z} }}\left(\bm Z, t\right) \delta f\left[\bm{\mathcal{Z} }\left(0;\bm Z, t\right),0\right]\right\rangle \right\},
\end{equation}
which is associated with the destruction term for the fluctuation dynamics, and its contribution to the evolution of averaged distribution is usually ignored by arguing that the initial perturbation can be ignored in statistical physics \cite{BalescuBook05}. Here we point out that dropping this $\Delta$ term in Eq. (\ref{eq:FP0}) is consistent with assumption (2) and assumption (1). The reason is as follows.

Comparing this term with the diffusion term, one has
\begin{equation}
\frac{\Delta}{\partial_{\bm Z}\cdot \bm D \cdot \partial_{\bm Z} F}\sim \frac{\delta f \lambda_c/\tau_c/L}{F \lambda_c^2/\tau_c/L^2}\sim \frac{\delta f L}{F \lambda_c}\sim \mathcal{O}\left(\epsilon\right),
\end{equation}
where assumption (2) on the scale separation was used, therefore, the $\Delta$ term in Eq. (\ref{eq:FP0}) shall be dropped.

To proceed, we prove the symmetry relation of the phase-space diffusion tensor.
Note that the Lagrangian integral in Eq. (\ref{eq:D0}) is taken along the particle trajectory which passes through $\bm Z$ at time $t$. The Eulerian integral in Eq. (\ref{eq:D0}) means that the relevant physical quantity should be averaged over a collection of trajectories, with each one labeled by $\left(\bm Z, t\right)$; a collection of these trajectories is essentially Ehrenfest's G-path \cite{TodaBook95}, as illustrated in Fig. 1. Using assumption (2), the up-end of this integral can be extended to $T/2$ (-T/2) when $t>0$ ($t>0$). Then following assumption (1), one can write
\begin{equation}
\bm D \left(\bm Z\right)=\frac{1}{2T}\int_{-\frac{T}{2}}^{\frac{T}{2}}d_E t \dot{\bm{\mathcal{Z} }}\left(\bm Z, t\right) \int_{-\frac{T}{2}}^{\frac{T}{2}}d_L t' \dot{\bm{\mathcal{Z} }}\left[\bm{\mathcal{Z} }\left(t';\bm Z, t\right), t'\right].\label{eq:D1}
\end{equation}
The integral in Eq. (\ref{eq:D1}) is a 2-dimensional one whose domain is illustrated by the shaded area (covered by the G-path) in Fig. 1. Changing the order of integration in Eq. (\ref{eq:D1}), one finds
\begin{equation}
D^{ij}=\frac{1}{2}\int_{-\infty}^{\infty}dt_0\left\langle \dot{\mathcal {Z}}^{i}\left( 0\right)\dot{\mathcal { Z}}^{j}\left(t_0\right)\right\rangle_{ens},
\end{equation}
or an alternative form for convenience,
\begin{equation}
D^{ij}=\int_{-\infty}^{0}dt_0\left\langle \dot{\mathcal{Z}}^{i}\left(0\right)\dot{\mathcal{Z}}^{j}\left(t_0\right)\right\rangle_{ens},
\end{equation}
where the limit $T/\tau_c\rightarrow\infty$ was taken.
The operator $\left\langle .\right\rangle_{ens}$ used above can be written in the following discrete form
\begin{equation}
\left\langle \mathcal A \left(0\right)\mathcal B \left(t_0\right)\right\rangle_{ens}
\equiv \lim_{N\rightarrow\infty}\frac{1}{N}\sum_{n=1}^{N}\mathcal{A} \left[\bm Z_n(0), 0\right]\mathcal{B} \left[\bm {Z}_n(t_0), t_0\right],\label{eq:Z-av}
\end{equation}
where $\mathcal B \left[\bm {Z}_n(t_0), t_0\right]$ means that $\mathcal B$ is evaluated at
\begin{equation}
\bm {Z}_n(t_0)=\bm{\mathcal{Z} }\left(t_0;\bm Z, t_n \right),\label{eq:G-path-mapping}
\end{equation}
the phase-space coordinate at time $t_0$ (open circles in Fig. 1) of the $n$th trajectory labeled by $\left(\bm Z, t_{n}\right)$ (black points in Fig. 1), with $t_{n+1}-t_{n}=T/N\ll \tau_c$.

Accordingly, one can write
\begin{equation}
\left\langle f \right\rangle_{ens}\left(\bm Z, t_0\right)
\equiv \lim_{N\rightarrow\infty}\frac{1}{N}\sum_{n=1}^{N} f \left[\bm{\mathcal{Z} }\left(t_0;\bm Z, t_n \right), t_0\right],\label{eq:Z-avf}
\end{equation}
where $t_n$'s (black points in Fig. 1) are uniformly distributed in the domain $\left(t-T/2, t+T/2\right)$, with $T\gg\tau_c$.

Clearly, the particle trajectory establishes a mapping [Eq. (\ref{eq:G-path-mapping})] between $t_{(n)}$ (black points in Fig. 1) and $\bm Z _{(n)}$ (open circles in Fig. 1), therefore, the $t-$ average $\left\langle . \right\rangle$ we take in this paper is equivalent to the $\bm Z -$ (ensemble) average $\left\langle .  \right\rangle_{ens}$ with its specific form given by Eq. (\ref{eq:Z-avf}), which is easily understood by analyzing Fig. 1.

Following assumption (1), one proves the symmetry relation of the phase-space diffusion tensor,
\begin{subequations}
\begin{eqnarray}
D^{ij}&\equiv&\frac{1}{2}\int_{-\infty}^{\infty}dt_0\left\langle \dot{\mathcal{Z}}^{i}\left(0\right)\dot{\mathcal{Z} }^{j}\left(t_0\right)\right\rangle\\
&=&\frac{1}{2}\int_{-\infty}^{\infty}dt_0\left\langle \dot{\mathcal{Z}}^{i}\left(0\right)\dot{\mathcal{ Z}}^{j}\left(-t_0\right)\right\rangle\\
&=&\frac{1}{2}\int_{-\infty}^{\infty}dt_0\left\langle \dot{\mathcal{Z}}^{i}\left(t_0\right)\dot{\mathcal{ Z}}^{j}\left(0\right)\right\rangle \equiv  D^{ji}.
\end{eqnarray}
\end{subequations}

To proceed, we prove the positive definiteness of the kinetic diffusion tensor.
The production rate of the entropy density
\begin{equation}
s \equiv -\int d^3\bm v F\ln F,
\end{equation}
is given by a quadratic form,
\begin{equation}
\left(\partial_t s\right)_{tur}=\int d^3\bm v F \partial_{i}\ln F D^{ij}\partial_j \ln F,\label{eq:ent-0}
\end{equation}
with the inhomogeneity of the averaged distribution denoted by $\partial_j \ln F$.
If $D^{ij}=0$, $\left(\partial_t s\right)_{tur}=0$; it conflicts with assumption (4). Therefore, one concludes that
\begin{equation}
D^{ij}\neq 0,
\end{equation}
which indicates the existence of the correlation is a natural result of the entropy principle; therefore, using the fact that the auto-correlation is always positive-definite \cite{PathriaBook12}, one finds
\begin{equation}
D^{ii}> 0.
\end{equation}

Note that Eq. (\ref{eq:ent-0}) can be written as
\begin{equation}
\left(\partial_t s\right)_{tur}=\int d^3\bm v F \int _{-\infty}^{\infty} dt_0 \frac{1}{2} \left\langle \mathcal{A}\left(0\right)\mathcal{A}\left(t_0\right)\right\rangle,\label{eq:ent-1}
\end{equation}
with $\mathcal{A}\left(t\right)=\dot {\mathcal{Z}^j}\left(t\right)\partial_j \ln F $; therefore, one concludes that $\left(\partial_t s\right)_{tur}>0$ is guaranteed by the positive-definiteness of both $F$ and the auto-correlation.

\begin{figure}
  \includegraphics[width=8.5cm]{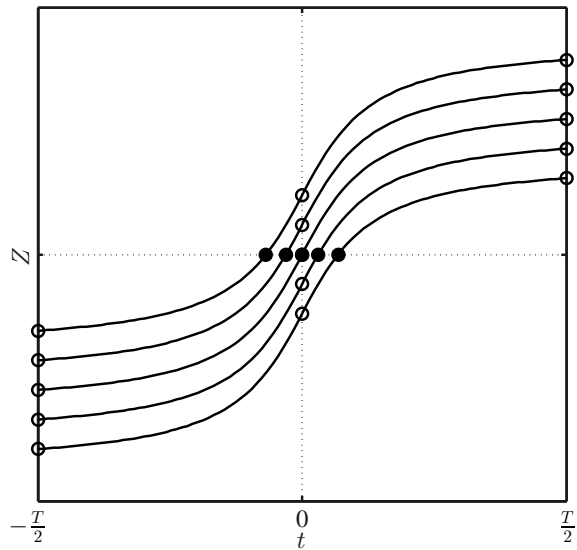}\\
  \caption{Illustration of the G-path (a collection of trajectories labeled by the black points), the $t-$ average (black points, $\bm Z$ fixed), and the $\bm Z-$ (ensemble) average (open circles, $t$ fixed). For a stochastic system, the trajectory may be similar to the orbit of a Brownian particle, and the open circles should be randomly distributed along the vertical line. A trajectory provides a mapping between the black point and the open circles. The G-path mapping ensures that the time average is equivalent to the ensemble average. } \label{fig1}
\end{figure}

The Fokker-Planck equation,
\begin{equation}
\partial_t F\left(\bm Z, t\right)+\partial_{\bm Z}\cdot \left[\left\langle\dot{\bm{\mathcal{Z} }}\right\rangle F\left(\bm Z, t\right) - \bm D\cdot \partial_{\bm Z} F\left(\bm Z, t\right)\right]=0,\label{eq:FP}
\end{equation}
which is thus derived from the first-principle, can be used to describe the nonlinear turbulent dissipation.

Although the quasilinear transport theory \cite{ChenJGR99,WangPoP12,BalescuBook05} can also gives a Fokker-Planck equation with a symmetric kinetic diffusion tensor obtained by integrating along the unperturbed orbit, it is based on the perturbation theory; therefore, the quasilinear theory is not applicable in the nonlinear turbulence saturation stage. Note that the diffusion tensor, Eq. (\ref{eq:D0} or \ref{eq:D1}), used in the Fokker-Planck equation derived here [Eq. (\ref{eq:FP})], is obtained by integrating along the full orbit; therefore, it does not depend on the perturbation method and applies in the nonlinear regime. It is straightforward to verify that the kinetic diffusion coefficient defined here agrees with the previous results in the limits of the random-walk model and the quasilinear transport model.

\section{Application}
If one assumes a Maxwellian averaged distribution,
\begin{equation}
F_M=\frac{n_0}{\left(2\pi T_0/m\right)^{3/2}}\exp\left(-\frac{K}{T_0}\right),
\end{equation}
with $K=mv^2/2$ the kinetic energy of particle; $n_0\left(x\right)$ and $T_0\left(x\right)$ the averaged density and temperature, respectively. Taking the velocity moment of Eq. (\ref{eq:FP}), one can find the particle flux and the heat flux. The detailed derivation is similar to Refs. \onlinecite{ChenJGR99,WangPoP12}, and the main results are summarized as follows.
\begin{subequations}
\begin{eqnarray}
\frac{\Gamma^x}{n_0}&=&- L_{11}\partial_x \ln p_0- L_{12}\partial_x \ln T_0+eE^{x}L_{11}\frac{1}{T_0},\label{eq:S-E}\\
\frac{q^x}{p_0}&=&- L_{21}\partial_x \ln p_0- L_{22}\partial_x \ln T_0+eE^{x}L_{21}\frac{1}{T_0},
\end{eqnarray}
\end{subequations}
with $\Gamma^x$ and $q^x$ the particle flux and heat flux, respectively; $p_0=n_0 \left(x\right)T_0 \left(x\right)$ the equilibrium pressure; $E^{x}$ the mean electric field. The transport matrix is given by
\begin{equation}
L_{ij}=\frac{1}{n_0}\int d^3 \bm v F_M D^{ x x}\left(\frac{K}{T_0}-\frac{5}{2}\right)^{i+j-2},
\end{equation}
which clearly satisfies Onsager's reciprocal relation \cite{OnsagerPR31a,OnsagerPR31b},
\begin{equation}
 L_{ij}=L_{ji}.
\end{equation}

The last term in Eq. (\ref{eq:S-E}) is related to the celebrated Stokes-Einstein relation \cite{PathriaBook12}. It is due to the kinetic diffusion coefficient
\begin{equation}
D^{xK}=eE^{x}D^{xx},
\end{equation}
which reflects a cross-correlation induced by the mean electric field. This term indicates that a perpendicular mean electric field in a turbulent magnetized plasma should drive a perpendicular particle flux. Note that this assertion does not depend on the details of the turbulence. The point is that $\delta K=eE^x \delta x$, with $\delta K$ and $\delta x$ the fluctuations of particle kinetic energy and position, respectively. Note that $-\partial_{K}\ln F =1/T_0$.

\section{Summary and discussions}
In summary, we have shown a first-principle derivation of the Fokker-Planck equation with a symmetric diffusion tensor from the nonlinear Vlasov equation, by introducing 4 physical assumptions. Assumption (1) restricts our discussion within the scope of stationary-homogeneous turbulence, as is widely adopted in fluid turbulence \cite{FrischBook95} and plasma turbulence \cite{BalescuBook05}. Assumption (2) is essentially the usual transport ansatz \cite{LifshitzBOOK81} based on which a local transport theory is established due to the separation of scales. Assumption (3) states that the ensemble average, whose classical definition makes it difficult to tract mathematically, can be evaluated by taking the time-average, which is a fundamental concept in statistical mechanics; we have discussed the equivalence of the $t-$ average and the $\bm Z-$ (ensemble) average based on the G-path mapping [Eq. (\ref{eq:G-path-mapping})] shown in Fig. 1 and the specific discrete form of the $\bm Z-$ average in Eq. (\ref{eq:Z-avf}); note that in usual statistical mechanics \cite{LifshitzBOOK81,PathriaBook12,TodaBook95} the $t-$ average is asserted to be equivalent to the ensemble ($\bm Z-$) average by introducing the ergodicity hypothesis, however, there is not a mathematically tractable form of the $\bm Z-$ average given out. Assumption (4) is the most important physical point of the deduction, which demonstrates that the Fokker-Planck equation is a natural result of the nonlinear Vlasov equation in view of the entropy principle. As an application of the present theory, we have shown that the Onsager symmetry relation and the general Stokes-Einstein relation hold for a magnetic confinement plasma in a general stationary-homogeneous nonlinear turbulence; in particular, a perpendicular mean electric filed can drive a perpendicular particle flux through the Stokes-Einstein relation, independent of the details of the assumed stationary-homogeneous turbulence, be it electrostatic or electromagnetic, ion-temperature-gradient driven or electron-temperature-gradient driven. Further discussions on the Stokes-Einstein relation and the perpendicular electric field shall be left for future publications.

To end this paper, we point out that the large-scale massive computer gyro-kinetic Vlasov-Poisson simulation has become one of the most powerful tools in investigating the nonlinear turbulence in plasmas, these tools include the particle-in-cell scheme \cite{LinScience98} and the semi-Lagrangian continuum scheme \cite{ChengJCP76,SonnendruckerJCP99}. The transport fluxes, such as the particle flux and the heat flux, can be found from the simulation data by computing the moments of numerical distribution function; however, the full transport matrix including the Onsager relation and the Stokes-Einstein relation predicted by the conventional transport theory based on various reduced models of turbulence \cite{BalescuBook05}, can hardly be obtained from these simulation data. Clearly, there is a serious lack of bridge between the nonlinear turbulence simulation and the transport theory. Dupree has introduced the renormalization method in deriving a Fokker-Planck equation to describe the slow-evolution of the averaged distribution function \cite{DupreePF66}. Detailed discussions relating Dupree's perturbation theory to the direct-interaction-approximation method can be found in Chap. 8.11 and Chap. 9.1 of Ref. \onlinecite{BalescuBook05}; briefly, this method is based on "a set of equations for the two-point two-time correlation function" "that is beyond any possibilities of solution, be it analytical or numerical". The present work may suggest a step toward solving this outstanding problem in understanding the nonlinear turbulent transport, since all the macroscopic transport properties, such as the Onsager relation and the Stokes-Einstein relation, are represented by the kinetic diffusion tensor, $D^{ij}$, which can be evaluated by using Eq. (\ref{eq:D0}) or Eq. (\ref{eq:D1}); the ensemble average in these equations can be carried out in a straightforward way in nonlinear turbulence simulation, which only involves evaluating a 2-dimensional integral, a numerically tractable problem. The key point is that the previous two-point ($\bm Z$ in phase space) two-time correlation problem is reduced here to a one-point ($\bm Z$) two-time correlation problem, which is accomplished by introducing the G-path mapping [Eq. (\ref{eq:G-path-mapping})].
The method proposed here may also be useful in studying other fluctuating systems in statistical physics, turbulence, biologies, and financial system.

\begin{acknowledgments}
This work was supported by the National Natural Science Foundation of China under Grant No. 11175178, No. 11375196 and the National ITER program of China under Contract No. 2014GB113000.
\end{acknowledgments}


\end{document}